\documentclass[11pt]{article}

\usepackage{fullpage}
\usepackage{latexsym}
\usepackage{amsfonts,amsmath}

\def\01{\{0,1\}}

\newcommand{\ket}[1]{|#1\rangle} 
\newcommand{\bra}[1]{\langle#1|} 
\newcommand{\inp}[2]{\langle{#1}|{#2}\rangle}

\newcommand{\DISJ}{\mbox{\rm DISJ}}
\newcommand{\INT}{\mbox{\rm INT}}
\newcommand{\EQ}{\mbox{\rm EQ}}
\newcommand{\NEQ}{\mbox{\rm NEQ}}

\newtheorem{theorem}{Theorem}
\newtheorem{lemma}{Lemma}

\newtheorem{corollary}{Corollary}

\newenvironment{proof}
{\noindent {\bf Proof. }}
{{\hfill $\Box$}\\
 \smallskip}

\begin{document}

\title{Improved Quantum Communication Complexity Bounds 
for Disjointness and Equality}
\author{Peter H{\o}yer\thanks{Dept.~of Computer Science, 
University of Calgary, Alberta, Canada T2N~1N4.
Email:~\mbox{\texttt{hoyer}\textbf{\char"40}\texttt{cpsc.ucalgary.ca}.}}
\and 
Ronald de Wolf\thanks{CWI, P.O.Box~94079, NL-1090 GB Amsterdam, The Netherlands.
Email: \mbox{\texttt{rdewolf}\textbf{\char"40}\texttt{cwi.nl}}.
Partially supported by the EU fifth framework project QAIP, IST--1999--11234,
and by TALENT grant S 62-565 from the Netherlands Organization 
for Scientific Research (NWO).}}
\date{}
\maketitle

\begin{abstract}
We prove new bounds on the quantum communication complexity
of the disjointness and equality problems.
For the case of exact and non-deterministic protocols
we show that these complexities are all equal to $n+1$, the previous best 
lower bound being $n/2$. We show this by improving a general bound for 
non-deterministic protocols of de~Wolf. 
We also give an $O(\sqrt{n}\cdot c^{\log^* n})$-qubit bounded-error protocol
for disjointness, modifying and improving the earlier $O(\sqrt{n}\log n)$ 
protocol of Buhrman, Cleve, and Wigderson,
and prove an $\Omega(\sqrt{n})$ lower bound for a large class of 
protocols that includes the BCW-protocol as well as our new protocol.\\[1mm]
\end{abstract}

\section{Introduction}

The area of {\em communication complexity\/} deals with abstracted
models of distributed computing, where one only cares about minimizing
the amount of communication between the parties and not about
the amount of computation done by the individual parties.
The standard setting is the following.
Two parties, Alice and Bob, want to compute some function 
$f:\01^n\times\01^n\rightarrow\01$.
Alice receives input $x\in\01^n$, Bob receives $y\in\01^n$, and they want 
to compute $f(x,y)$. For example, they may want to find out whether
$x=y$ (the {\em equality\/} problem) or whether $x$ and $y$ are
characteristic vectors of disjoint sets (the {\em disjointness\/} problem).
A communication protocol is a distributed algorithm where Alice first does 
some computation on her side, then sends a message to Bob, who does some 
computation on his side, sends a message back, etc. 
The {\em cost} of the protocol is measured by the number of bits 
(or qubits, in the quantum case) communicated on the worst-case input $(x,y)$.

As in many other branches of complexity theory, we can distinguish
between various different ``modes'' of computation.
Letting $P(x,y)$ denote the acceptance probability of the protocol
(the probability of outputting 1), we will consider 4 different kinds 
of protocols for computing $f$:
\begin{itemize}
\item An {\em exact} protocol has $P(x,y)=f(x,y)$, for all $x,y$
\item A {\em non-deterministic} protocol has $P(x,y)>0$ if and only 
if $f(x,y)=1$, 
for all $x,y$
\item A {\em one-sided error} protocol has $P(x,y)\geq 1/2$ if $f(x,y)=1$, 
and $P(x,y)=0$ if $f(x,y)=0$
\item A {\em two-sided error} protocol has $|P(x,y)-f(x,y)|\leq 1/3$, 
for all $x,y$
\end{itemize}
These 4 modes of computation correspond to those of 
the computational complexity classes P, NP, RP, and BPP, respectively.

Protocols may be {\em classical} (send and process classical bits)
or {\em quantum} (send and process quantum bits).
Classical communication complexity was introduced by Yao~\cite{yao:distributive},
and has been studied extensively. It is well-motivated by its intrinsic
interest as well as by its applications in lower bounds on circuits,
VLSI, data structures, etc.  We refer to the book of 
Kushilevitz and Nisan~\cite{kushilevitz&nisan:cc} for definitions and results.
We will use $D(f)$, $N(f)$, $R_1(f)$, and $R_2(f)$ to denote the
minimal cost of classical protocols for $f$ in the exact, non-deterministic,
one-sided error, and two-sided error settings, 
respectively.\footnote{Kushilevitz and
Nisan~\cite{kushilevitz&nisan:cc} use $N^1(f)$ for our $N(f)$, $R^1(f)$ for 
our $R_1(f)$ and $R(f)$ for our $R_2(f)$.}
Note that $R_2(f)\leq R_1(f)\leq D(f)\leq n+1$ and
$N(f)\leq R_1(f)\leq D(f)\leq n+1$ for all $f$.
Similarly we define $Q_E(f)$, $NQ(f)$, $Q_1(f)$, and $Q_2(f)$ for the
quantum versions of these communication complexities (we will be a bit more 
precise about the notion of a quantum protocol in the next section).
For these complexities, we assume Alice and Bob start out without
any shared randomness or entanglement.

Quantum communication complexity was introduced by (again) Yao~\cite{yao:qcircuit}
and the first examples of functions where quantum communication complexity
is less than classical communication complexity were given 
in~\cite{cleve&buhrman:subs,bdht:multiparty,cdnt:ip,BuhrmanCleveWigderson98}.
In particular, Buhrman, Cleve, and Wigderson~\cite{BuhrmanCleveWigderson98} 
showed for a specific {\em promise version} of the equality problem that  
$Q_E(f)\in O(\log n)$ while $D(f)\in\Omega(n)$. 
They also showed for the {\em intersection\/} problem
(the negation of the disjointness problem) that $Q_1(\INT_n)\in O(\sqrt{n}\log n)$,
whereas $R_2(\INT_n)\in\Omega(n)$ is a well known and non-trivial result
from classical communication complexity~\cite{ks:disj,razborov:disj}.
Later, Raz~\cite{raz:qcc} exhibited a promise problem with an exponential 
quantum-classical gap even in the bounded-error setting: 
$Q_2(f)\in O(\log n)$ versus $R_2(f)\in\Omega(n^{1/4}/\log n)$.
Other results on quantum communication complexity may be found in
\cite{kremer:thesis,cdnt:ip,astv:qsampling,nielsen:thesis,buhrman&wolf:qcclower,klauck:qpcom,wolf:ndetq,kntz:qinteraction,klauck:qcclower}.

The aim of this paper is to sharpen the bounds on the quantum
communication complexities of the equality and disjointness (or
intersection) problems, in the 4 modes we distinguished above. 
We summarize what was known prior to this paper:
\begin{itemize}
\item $n/2\leq Q_1(\EQ_n),Q_E(\EQ_n)\leq n+1$~\cite{kremer:thesis,buhrman&wolf:qcclower}\\
$n/2\leq NQ(\EQ_n)\leq n+1$~\cite{wolf:ndetq}\\
$Q_2(\EQ_n)\in\Theta(\log n)$~\cite{kremer:thesis}
\item $n/2\leq Q_1(\DISJ_n),Q_E(\DISJ_n)\leq n+1$~\cite{kremer:thesis,buhrman&wolf:qcclower}\\
$n/2\leq NQ(\DISJ_n)\leq n+1$~\cite{wolf:ndetq}\\
$\log n\leq Q_1(\INT_n),Q_2(\DISJ_n)\in O(\sqrt{n}\log n)$~\cite{BuhrmanCleveWigderson98}
\end{itemize}
In Section~\ref{secndet} we first sharpen the non-deterministic bounds,
by proving a general algebraic characterization of $NQ(f)$.
In~\cite{wolf:ndetq} it was shown for all functions $f$ that
$$
\frac{\log \textit{nrank}(f)}{2}\leq NQ(f)\leq \log(\textit{nrank}(f))+1,
$$ 
where $\textit{nrank}(f)$ denotes the rank of a ``non-deterministic matrix'' for $f$
(to be defined more precisely below). 
It is interesting to note that in many places in quantum computing one 
sees factors of $\frac{1}{2}$ appearing that are essential, for example
in the query complexity of parity~\cite{bbcmw:polynomials,fggs:parity},
in the bounded-error query complexity of all functions~\cite{dam:oracle},
in superdense coding~\cite{superdense}, and in lower bounds for 
entanglement-enhanced quantum communication 
complexity~\cite{buhrman&wolf:qcclower,nielsen:thesis}.
In contrast, we show here that the $\frac{1}{2}$ in the above lower
bound can be dispensed with, and the upper bound is tight:%
\footnote{Similarly we can improve the query complexity result
$ndeg(f)/2\leq NQ_q(f)\leq ndeg(f)$ of~\cite{wolf:ndetq} to 
the optimal $NQ_q(f)=ndeg(f)$.}
$$
NQ(f)=\log(\textit{nrank}(f))+1.
$$ 
Equality and disjointness both have non-deterministic rank $2^n$,
so their non-deterministic complexities are maximal: $NQ(\EQ_n)=NQ(\DISJ_n)=n+1$.
(This contrasts with their complements: 
$NQ(\NEQ_n)=2$~\cite{mbcc:siment} and $NQ(\INT_n)\leq N(\INT_n)=\log n+1$.)
Since $NQ(f)$ lower bounds $Q_1(f)$ and $Q_E(f)$, we also obtain optimal bounds
for the one-sided and exact quantum communication complexities of
equality and disjointness. In particular, $Q_E(\EQ_n)=n+1$, which answers
a question posed to one of us (RdW) by Gilles Brassard in December 2000.

The two-sided error bound $Q_2(\EQ_n)\in\Theta(\log n)$ is easy to show,
whereas the two-sided error complexity of disjointness is still wide open.
In Section~\ref{secdisj} we give a one-sided error 
protocol for the intersection problem that improves the 
$O(\sqrt{n}\log n)$ protocol of Buhrman, Cleve, and Wigderson
by nearly a log-factor:
$$
Q_1(\INT_n)\in O(\sqrt{n}\cdot c^{\log^\star n}),
$$
where $c$ is a (small) constant.
The function $\log^\star n$ is defined as the minimum number of
iterated applications of the logarithm function necessary to obtain a
number less than or equal to~1: 
$\log^\star n = \min\{r \geq 0\mid \log^{(r)}n \leq 1\}$, where
$\log^{(0)}$ is the identity function and 
$\log^{(r)} = \log \circ \log^{(r-1)}$.
Even though $c^{\log^\star n}$ is exponential in~$\log^\star n$, 
it is still very small in~$n$, in particular $c^{\log^\star n} \in
o(\log^{(r)} n)$ for every constant~$r \geq 1$.
It should be noted that our protocol is asymptotically somewhat more efficient
than the BCW-protocol ($\sqrt{n}c^{\log^\star n}$ versus $\sqrt{n}\log n$), 
but is also more complicated to describe; it is based on a recursive 
modification of the BCW-protocol, an idea that has also been used for 
claw-finding by Buhrman \emph{et~al.}~\cite[Section~5]{betal:distinctness}.

Proving good \emph{lower} bounds on the $Q_2$-complexity of the disjointness
and intersection problems is one of the main open problems in quantum
communication complexity. Only logarithmic lower bounds are known so far
for general protocols~\cite{kremer:thesis,astv:qsampling,buhrman&wolf:qcclower}.
The lower bound $\Omega(n^{1/k}/k^3)$ was shown in~\cite{kntz:qinteraction} 
for protocols exchanging at most $k$ messages.
In Section~\ref{ssecdisjlower} we prove a nearly tight lower bound 
of $\Omega(\sqrt{n})$ qubits of communication for all  
protocols that satisfy the constraint that their acceptance probability 
is a function of $x\wedge y$ (the $n$-bit AND of 
Alice's $x$ and Bob's $y$), rather than of $x$ and $y$ ``separately''. 
Since $\DISJ_n$ itself is also a function only of $x\wedge y$,
this does not seem to be an extremely strong constraint.
The constraint is satisfied by a large class of natural protocols, 
in particular by the BCW-protocol and by our new protocol.
It seems plausible that the general bound is $Q_2(\DISJ_n)\in\Omega(\sqrt{n})$
as well, but we have so far not been able to weaken the constraint that 
the acceptance probability is a function of $x\wedge y$.

\section{Preliminaries}

\subsection{Quantum computing}

Here we briefly sketch the setting of quantum computation,
referring to the book of Nielsen and Chuang~\cite{nielsen&chuang:qc} 
for more details. An $m$-qubit quantum {\em state} $\ket{\phi}$ is a 
superposition or linear combination over all classical $m$-bit states:
$$
\ket{\phi}=\sum_{i\in\01^m}\alpha_i\ket{i},
$$
with the constraint that $\sum_i|\alpha_i|^2=1$.
Equivalently, $\ket{\phi}$ is a unit vector in $\mathbb{C}^{2^m}$.
Quantum mechanics allow us to change this state by means of
unitary (i.e., norm-preserving) operations: 
$\ket{\phi_{\text{new}}}=U\ket{\phi}$,
where $U$ is a $2^m\times 2^m$ unitary matrix. 
A {\em measurement} of $\ket{\phi}$ produces the outcome $i$ 
with probability $|\alpha_i|^2$, and then leaves the system in 
the state $\ket{i}$.

The two main examples of quantum algorithms so far, are Shor's algorithm 
for factoring $n$-bit numbers using $\text{poly}(n)$ elementary unitary 
transformations~\cite{shor:factoring} and Grover's algorithm 
for searching an unordered $n$-element space using $O(\sqrt{n})$ 
``look-ups'' or queries in the space~\cite{grover:search}.
Below we use a technique called {\em amplitude amplification},
which generalizes Grover's algorithm:

\begin{theorem}[Amplitude amplification~\cite{bhmt:countingj}]\label{thm:amplampl}
There exists a quantum algorithm \textup{\textbf{QSearch}}
with the following property. Let $\mathcal A$ be any quantum algorithm 
that uses no measurements, and let
$\chi:\{1,\ldots,n\}\rightarrow\{0,1\}$ be any Boolean function.
Let $a$ denote the initial success probability of~$\mathcal A$ 
of finding a solution (i.e., the probability of 
outputting some~$z\in\{1,\ldots,n\}$ so that $\chi(z)=1$).
Algorithm \textup{\textbf{QSearch}} finds a solution using 
an expected number of $O\left(\frac{1}{\sqrt{a}}\right)$ applications 
of $\mathcal{A}$, $\mathcal{A}^{-1}$, and $\chi$ if $a>0$, 
and it runs forever if $a=0$.
\end{theorem}

Consider the problem of searching an unordered $n$-element space.
An algorithm $\mathcal A$ that creates a uniform superposition over 
all $i\in\{1,\ldots,n\}$ has success probability $a\geq 1/n$, so plugging
this into the above theorem and terminating after $O(\sqrt{n})$
applications gives us an algorithm that finds a solution 
with probability $\geq 1/2$ provided there is one, and otherwise
outputs `no solution'.

\subsection{Communication complexity}

For classical communication protocols we refer to~\cite{kushilevitz&nisan:cc}.
Here we briefly define quantum communication protocols, referring to the
surveys~\cite{tashma:qcc,buhrman:qccsurvey,klauck:qccsurvey,brassard:qcc}
for more details.

The space in which the quantum protocol works consists of 3 parts:
Alice's part, the communication channel, and Bob's part (we will not
write the dimensions of these spaces explicitly).
Initially these 3 parts contain only 0-qubits:
$$
\ket{0}\ket{0}\ket{0}.
$$
We assume Alice starts the protocol.
Alice applies a unitary transformation $U^A_1(x)$ to her part and the 
channel. This corresponds to her initial computation and her first message.
The length of this message is the number of channel qubits affected.
The state is now
$$
(U^A_1(x)\otimes I^B)\ket{0}\ket{0}\ket{0},
$$
where $\otimes$ denotes tensor product, and $I^B$ denotes 
the identity transformation on Bob's part. 
Then Bob applies a unitary transformation $U^B_2(y)$ to his part and the channel.
This operation corresponds to Bob's reading Alice's message, doing some computation,
and putting a return-message on the channel. This process goes back and forth for 
some $k$ messages, so the final state of the protocol on input $(x,y)$ will be 
(in case Alice goes last)
$$
(U^A_k(x)\otimes I^B)(I^A\otimes U^B_{k-1}(y))\cdots
(I^A\otimes U^B_2(y))(U^A_1(x)\otimes I^B)\ket{0}\ket{0}\ket{0}.
$$
The total {\em cost} of the protocol is the total length 
of all messages sent, on the worst-case input $(x,y)$. 
For technical convenience, we assume that at the end of the protocol the 
output bit is the first qubit on the channel. 
Thus the acceptance probability $P(x,y)$ of the protocol is the probability that 
a measurement of the final state gives a `1' in the first channel-qubit.
Note that we do not allow intermediate measurements during the protocol.
This is without loss of generality: it is well known that such measurements
can be postponed until the end of the protocol at no extra communication cost.
As mentioned in the introduction, we use $Q_E(f)$, $NQ(f)$, $Q_1(f)$, and $Q_2(f)$ to
denote the cost of optimal exact, non-deterministic, one-sided error,
and two-sided error protocols for $f$, respectively.

The following lemma was stated summarily without proof by Yao~\cite{yao:qcircuit} 
and in more detail by Kremer~\cite{kremer:thesis}. It is key to many of 
the earlier lower bounds on quantum communication complexity as well 
as to ours, and is easily proven by induction on $\ell$.

\begin{lemma}[Yao~\cite{yao:qcircuit}; Kremer~\cite{kremer:thesis}]\label{lemkremer}
The final state of an $\ell$-qubit protocol on input $(x,y)$ can be written as
$$
\sum_{i\in\01^\ell}\ket{A_i(x)}\ket{i_\ell}\ket{B_i(y)},
$$
where the $A_i(x),B_i(y)$ are vectors (not necessarily of norm 1), 
and $i_\ell$ denotes the last bit of the $\ell$-bit string $i$ (the output bit).
\end{lemma}

The acceptance probability $P(x,y)$ of the protocol is 
the squared norm of the part of the final state that has $i_{\ell}=1$.
Letting $M_{ij}$ be the $2^n\times 2^n$ matrix whose $x,y$-entry 
is the inner product $\inp{A_i(x)}{B_j(y)}$, we can write $P$ (viewed as 
a $2^n\times 2^n$ matrix) as the sum $\sum_{i,j:i_{\ell}=j_{\ell}=1}M_{ij}$ of 
$2^{2\ell-2}$ rank 1 matrices, so the rank of $P$ is $\leq 2^{2\ell-2}$.
For example, for exact protocols this gives immediately that
$\ell$ is lower bounded by $\frac{1}{2}$ times the log of the rank of 
the communication matrix, and for non-deterministic protocols $\ell$ is 
lower bounded by $\frac{1}{2}$ times the log of the non-deterministic rank.
In the next section we will show how we can get rid of the factor 
$\frac{1}{2}$ in the non-deterministic case.

We use $x\wedge y$ for the $n$-bit string obtained by bitwise-ANDing 
$x$ and $y$, and similarly $x\oplus y$ for XOR.
Let $OR$ denote the $n$-bit function which is 1 if at least one of its 
$n$ input bits is 1, and $NOR$ be its negation.
We will be concerned with the following communication 
complexity problems:
\begin{itemize}
\item {\em Equality}: $\EQ_n(x,y)=NOR(x\oplus y)$
\item {\em Intersection}: $\INT_n(x,y)=OR(x\wedge y)$
\item {\em Disjointness}: $\DISJ_n(x,y)=NOR(x\wedge y)$
\end{itemize}

\section{Optimal non-deterministic bounds}\label{secndet}

Let $f:\01^n\times\01^n\rightarrow\01$.
A $2^n\times 2^n$ complex matrix $M$ is called a 
{\em non-deterministic matrix} for $f$ if it has the property
that $M_{xy}\neq 0$ if and only if $f(x,y)=1$
(equivalently, $M_{xy}=0$ if and only if $f(x,y)=0$).
We use $\textit{nrank}(f)$ to denote the {\em non-deterministic rank} of $f$,
which is the minimal rank among all non-deterministic 
matrices for $f$. In~\cite{wolf:ndetq} it was shown that
$$
\frac{\log \textit{nrank}(f)}{2}\leq NQ(f)\leq \log(\textit{nrank}(f))+1.
$$ 
In this section we show that the upper bound is the true bound.
The proof uses the following technical lemma.

\begin{lemma}\label{lemvectorsnrank}
If there exist sets $\{A_1(x),\ldots,A_m(x)\}\subseteq\mathbb{C}^d$
and $\{B_1(y),\ldots,B_m(y)\}\subseteq\mathbb{C}^d$ such that for all 
$x\in\01^n$ and $y\in\01^n$ we have:
$$
\sum_{i=1}^m A_i(x)\otimes B_i(y)= 0 \mbox{ if and only if } f(x,y)=0,
$$
then $\textit{nrank}(f)\leq m$.
\end{lemma}

\begin{proof}
We will use $A_i(x)_j$ to denote the $j$th entry of the vector $A_i(x)$.
We use pairs $(j,k)\in\{1,\ldots,d\}^2$ to index entries
of vectors in the $d^2$-dimensional tensor space. 
Note that
\begin{quote}
if $f(x,y)=0$ then $\sum_{i=1}^m A_i(x)_jB_i(y)_k=0$ for all $(j,k)$\\
if $f(x,y)=1$ then $\sum_{i=1}^m A_i(x)_jB_i(y)_k\neq 0$ for some $(j,k)$
\end{quote}
As a first step, we want to replace the vectors $A_i(x)$ and $B_i(y)$ 
by numbers $a_i(x)$ and $b_i(y)$ that have similar properties.
We will use the probabilistic method~\cite{alon&spencer:probmethod} 
to show that this can be done.

Let $I$\/ be an arbitrary set of $2^{2n+1}$ non-zero numbers.
Choose coefficients $\alpha_1,\ldots,\alpha_d$ and $\beta_1,\ldots,\beta_d$, 
each coefficient picked uniformly at random from $I$.
For every $x$, define
$a_i(x)=\sum_{j=1}^d\alpha_j A_i(x)_j$,
and for every $y$ define $b_i(y)=\sum_{k=1}^d\beta_k B_i(y)_k$.
Consider the number
$$
v(x,y)=\sum_{i=1}^m a_i(x)b_i(y)=
\sum_{j,k=1}^d\alpha_j\beta_k\left(\sum_{i=1}^m A_i(x)_jB_i(y)_k\right).
$$
If $f(x,y)=0$, then $v(x,y)=0$ for all choices of the $\alpha_j,\beta_k$.

Now consider some $(x,y)$ with $f(x,y)=1$. There is a 
$(j',k')$ for which $\sum_{i=1}^m A_i(x)_{j'}B_i(y)_{k'}\neq 0$.
We want to prove that $v(x,y)=0$ happens only with very small probability.
In order to do this, fix the random choices of all 
$\alpha_j$, $j\neq j'$, and $\beta_k$, $k\neq k'$,
and view $v(x,y)$ as a function of the two remaining not-yet-chosen
coefficients $\alpha=\alpha_{j'}$ and $\beta=\beta_{k'}$:
$$
v(x,y)=c_0\alpha\beta+c_1\alpha+c_2\beta+c_3.
$$
Here we know that 
$c_0=\sum_{i=1}^m A_i(x)_{j'}B_i(y)_{k'}\neq 0$.
There is at most one value of $\alpha$ for which $c_0\alpha+c_2=0$.
All other values of $\alpha$ turn $v(x,y)$ into a linear equation 
in $\beta$, so for those $\alpha$ there is at most one choice 
of $\beta$ that gives $v(x,y)=0$. Hence out of the $(2^{2n+1})^2$ 
different ways to choose $(\alpha,\beta)$, at most 
$2^{2n+1}+(2^{2n+1}-1)\cdot 1<2^{2n+2}$ choices give $v(x,y)=0$.
Therefore:
$$
\Pr[v(x,y)=0]<\frac{2^{2n+2}}{(2^{2n+1})^2}=2^{-2n}.
$$
Using the union bound, we now have
$$
\Pr\left[\mbox{there is an }(x,y)\in f^{-1}(1)\mbox{ for which }v(x,y)=0\right] \
\leq \sum_{(x,y)\in f^{-1}(1)}\Pr[v(x,y)=0]< 2^{2n}\cdot 2^{-2n}=1.
$$
This probability is strictly less than 1, so there 
exist sets $\{a_1(x),\ldots,a_m(x)\}$ and $\{b_1(y),\ldots,b_m(y)\}$ 
that make $v(x,y)\neq 0$ for every $(x,y)\in f^{-1}(1)$. 
We thus have:
$$
\sum_{i=1}^m a_i(x)b_i(y)=0 \mbox{ if and only if } f(x,y)=0.
$$
View the $a_i$ and $b_i$ as $2^n$-dimensional vectors, 
let $A$ be the $2^n\times m$ matrix having the $a_i$ as columns,
and $B$ be the $m\times 2^n$ matrix having the $b_i$ as rows.
Then $(AB)_{xy}=\sum_{i=1}^m a_i(x)b_i(y)$, which is 0 if and only 
if $f(x,y)=0$. 
Thus $AB$ is a non-deterministic matrix for $f$, 
and $\textit{nrank}(f)\leq rank(AB)\leq rank(A)\leq m$.
\end{proof}

This lemma allows us to prove tight bounds 
for non-deterministic quantum protocols:

\begin{theorem}\label{thNQ=nrank}
$NQ(f)=\log(\textit{nrank}(f))+1$.
\end{theorem}

\begin{proof}
The upper bound $NQ(f)\leq \log(\textit{nrank}(f))+1$ was shown 
in \cite{wolf:ndetq} (actually, the upper bound shown there was 
$\log(\textit{nrank}(f))$ for protocols where only Bob has to know 
the output value). 
For the sake of completeness we repeat that proof here.
Let $r=nrank(f)$ and $M$ be a rank-$r$
non-deterministic matrix for $f$. Let $M^T=U\Sigma V$ be the
singular value decomposition of the transpose of $M$~\cite{horn&johnson:ma},
so $U$ and $V$ are unitary, and $\Sigma$ is a diagonal matrix
whose first $r$ diagonal entries are positive real numbers and
whose other diagonal entries are 0.
Below we describe a one-round non-deterministic protocol for $f$,
using $\log(r)+1$ qubits.
First Alice prepares the state $\ket{\phi_x}=c_x\Sigma V\ket{x}$,
where $c_x>0$ is a normalizing real number that depends on $x$.
Because only the first $r$ diagonal entries of $\Sigma$ are non-zero,
only the first $r$ amplitudes of $\ket{\phi_x}$ are non-zero,
so $\ket{\phi_x}$ can be compressed into $\log r$ qubits.
Alice sends these qubits to Bob. Bob then applies $U$ to $\ket{\phi_x}$
and measures the resulting state.
If he observes $\ket{y}$ then he puts 1 on the channel and otherwise 
he puts 0 on the channel.
The acceptance probability of this protocol is
$$
P(x,y) = |\bra{y}U\ket{\phi_x}|^2
       = c_x^2|\bra{y}U\Sigma V\ket{x}|^2
       = c_x^2|M^T_{yx}|^2
       = c_x^2|M_{xy}|^2.
$$
Since $M_{xy}$ is non-zero if and only if $f(x,y)=1$, 
$P(x,y)$ will be positive if and only if $f(x,y)=1$.
Thus we have a non-deterministic quantum protocol for $f$ 
with $\log(r)+1$ qubits of communication.

For the lower bound, consider a non-deterministic $\ell$-qubit protocol for $f$.
By the Yao-Kremer lemma (Lemma~\ref{lemkremer}), 
its final state on input $(x,y)$ can be written as
$$
\sum_{i\in\01^\ell}\ket{A_i(x)}\ket{i_\ell}\ket{B_i(y)}.
$$
Without loss of generality we assume the vectors $A_i(x)$ and $B_i(y)$ 
all have the same dimension $d$. Let $S=\{i\in\01^\ell \mid i_{\ell}=1\}$ 
and consider the part of the state that corresponds to output 1
(we drop the $i_{\ell}=1$ and the $\ket{\cdot}$-notation here):
$$
\phi(x,y)=\sum_{i\in S}A_i(x)\otimes B_i(y).
$$
Because the protocol has acceptance probability 0 if and only if $f(x,y)=0$,
this vector $\phi(x,y)$ will be the zero vector if and only if $f(x,y)=0$.
Now the previous lemma gives $\textit{nrank}(f)\leq |S|=2^{\ell-1}$, 
hence we obtain $\log(\textit{nrank}(f))+1\leq NQ(f)$.
\end{proof}

Note that a non-deterministic matrix for the equality function
will have non-zeroes on its diagonal and zeroes off-diagonal,
and hence will have full rank.
Hence $NQ(\EQ_n)=n+1$, which contrasts sharply with the
non-deterministic complexity of its complement (inequality),
which is only 2~\cite{mbcc:siment}.
Similarly, a non-deterministic matrix for disjointness will have
full rank, because reordering the rows gives an upper triangular
matrix with non-zeroes on its diagonal.
This gives tight bounds for the exact, one-sided error,
and non-deterministic settings:

\begin{corollary}
$Q_E(\EQ_n)=Q_1(\EQ_n)=NQ(\EQ_n)=Q_E(\DISJ_n)=Q_1(\DISJ_n)=NQ(\DISJ_n)=n+1$.
\end{corollary}


\section{On the bounded-error complexity of disjointness}\label{secdisj}

\subsection{Improved upper bound}\label{ssecdisjlower}

Here we show that we can take off most of the $\log n$ factor from
the $O(\sqrt{n}\log n)$ protocol for the intersection problem that
was given in~\cite{BuhrmanCleveWigderson98}.

\begin{theorem}\label{thdisjupper}
$Q_1(\INT_n)\in O(\sqrt{n}\cdot c^{\log^\star n})$.
\end{theorem}

\begin{proof}
We will recursively build a one-sided error protocol that can find 
an index $i$ such that $x_i=y_i=1$, if such an $i$ exists (call such 
an $i$ a `solution'). Clearly this suffices for computing $\INT_n(x,y)$.
Let $C_n$ denote the cost of our protocol on $n$-bit inputs.

Alice and Bob divide the $n$ indices $\{1,\ldots,n\}$ into 
$n/(\log n)^2$ blocks of $(\log n)^2$ indices each.
Alice picks a random number $j\in\{1,\ldots,n/(\log n)^2\}$
and sends the number $j$ to Bob. Now they run our protocol on 
the $j$th block, at a cost of $C_{(\log n)^2}$ qubits of communication.
Alice now measures her part of the state, and they verify whether
the measured $i$ is indeed a solution.
If there was a solution in the $j$th block, then Alice finds 
it with probability $\geq 1/2$, so the overall probability 
of finding a solution (if there is one) is $\geq (\log n)^2/2n$.
By using a superposition over all $j$ we can push all intermediate
measurements to the end without affecting the success probability.
Therefore, applying $O(\sqrt{n}/\log n)$ rounds of amplitude amplification 
(Theorem~\ref{thm:amplampl}) boosts this protocol to error $\leq 1/2$.
We thus have the recursion
$$
C_n\leq O(1)\frac{\sqrt{n}}{\log n}\left(C_{(\log n)^2}+O(\log n)\right).
$$
Since $C_1=2$, this recursion unfolds to the bound
$C_n\in O(\sqrt{n}\cdot c^{\log^\star n})$ for some constant $c$.
\end{proof}

\subsection{Lower bound for a large class of protocols}

Now we show a lower bound for two-sided error quantum protocols
for disjointness. The lower bound applies to all protocols
whose acceptance probability $P(x,y)$ is a function just
of $x\wedge y$, rather than of $x$ and $y$ ``separately''.
In particular, the protocols of~\cite{BuhrmanCleveWigderson98}
and of our previous section fall in this class. 

The lower bound basically follows by combining  
various results from~\cite{buhrman&wolf:qcclower}:

\begin{theorem}\label{thdisjlower}
Any two-sided error quantum protocol for $\DISJ_n$ 
whose acceptance probability is a function
of $x\wedge y$, has to communicate $\Omega(\sqrt{n})$ qubits.
\end{theorem}

\begin{proof}
Consider an $\ell$-qubit protocol with error $\leq 1/3$.  
By the comment following Lemma~\ref{lemkremer}, we can write 
its acceptance probability $P(x,y)$ as a $2^n\times 2^n$ 
matrix of rank $r\leq 2^{2{\ell}-2}$.
By \cite[Lemma~2]{buhrman&wolf:qcclower}, we can write
$P(x,y)=\sum_{i=1}^r a_i(x)b_i(y)$, where the $a_i$ and $b_i$ are $n$-variate 
multilinear polynomials.  Multiplying this out gives a lot of monomials
of $x$ and $y$ variables.  Because $P(x,y)$ is a function of $x\wedge y$, 
it only contains ``even'' monomials (i.e., where the set of $x$ and 
$y$-variables is the same, like $x_1x_3y_1y_3$, but not $x_1y_1y_3$ 
or $x_1x_2$).  For such polynomials, \cite[Lemma~3]{buhrman&wolf:qcclower} 
implies that the number of monomials in $P(x,y)$ equals $r$.  
Now identifying $x_i$ and $y_i$ in $P(x,y)$ gives an approximating 
polynomial for the $n$-bit function $NOR$, because $P(x,y)$ approximates 
$\DISJ_n(x,y)=NOR(x\wedge y)$. 
But \cite[Theorem~8]{buhrman&wolf:qcclower} implies
that such a polynomial has at least $2^{\sqrt{n/12}}$ monomials.
Hence $2^{\sqrt{n/12}}\leq r \leq 2^{2\ell-2}$, which gives
$\ell\geq\sqrt{n/48}+1$.
\end{proof}

\section{Open problems}

This paper fits in a sequence of papers that (slowly) extend what 
is known for quantum communication complexity, e.g.,
\cite{BuhrmanCleveWigderson98,astv:qsampling,raz:qcc,buhrman&wolf:qcclower,klauck:qpcom,wolf:ndetq,kntz:qinteraction,klauck:qcclower}.
The main open question is still the bounded-error
complexity of disjointness. 
Of interest is whether it is possible to prove an $O(\sqrt{n})$ upper
bound for disjointness, thus getting rid of the factor 
of $c^{\log^\star n}$ in our upper bound of Theorem~\ref{thdisjupper},
and whether it is possible to extend the lower bound of 
Theorem~\ref{thdisjlower} to broader classes of protocols.
Since disjointness is coNP-complete for communication complexity
problems~\cite{bfs:classes}, strong lower bounds on the disjointness 
problem imply a host of other lower bounds.

A second question is whether qubit communication can be significantly reduced 
in case Alice and Bob can make use of prior entanglement (shared EPR-pairs). 
Giving Alice and Bob $n$ shared EPR-pairs trivializes the non-deterministic 
complexity (use the EPR-pairs as a public coin to randomly guess 
some $n$-bit $z$, Alice then sends Bob 1 bit indicating whether $x=z$,
if $x=z$ then Bob can compute the answer $f(x,y)$ and send it to Alice, 
if $x\neq z$ then they output 0), but for the exact and bounded-error models 
it is open whether prior entanglement can make a significant difference.

\subsection*{Acknowledgments}
We thank Harry Buhrman and Hartmut Klauck for helpful discussions
concerning the proof of Lemma~\ref{lemvectorsnrank}.


\end{document}